\documentclass{article}
\usepackage[utf8]{inputenc}
\usepackage{amsmath}
\usepackage{amsthm}
\newtheorem{prop}{Proposition}
\newtheorem{theorem}{Theorem}

\usepackage{enumerate}   
\usepackage[square, numbers]{natbib}
\usepackage[affil-it]{authblk}
\usepackage{hyperref}
\providecommand{\keywords}[1]
{
  \small	
  \textbf{\textit{Keywords---}} #1
}

\title{A min-max theorem for the \\minimum fleet-size problem}

\author[a]{Tinghan (Joe) Ye}
\author[a]{David Shmoys \footnote{ \noindent Corresponding author. \\ \indent E-mail addresses: \href{mailto:ty357@cornell.edu}{ty357@cornell.edu}(T. Ye), \href{mailto:david.shmoys@cornell.edu}{david.shmoys@cornell.edu}(D. Shmoys) \\ \indent Postal address: 231 Rhodes Hall, 136 Hoy Road, Cornell University, Ithaca, NY 14853
}}
\affil[a]{School of Operations Research and Information Engineering, Cornell University}
\date{}

\begin{document}

\maketitle

\begin{abstract}
    A retrospective fleet-sizing problem can be solved via bipartite matching, where a maximum cardinality matching corresponds to the minimum number of vehicles needed to cover all trips. We prove a min-max theorem on this minimum fleet-size problem: the maximum number of pairwise incompatible trips is equal to the minimum fleet size needed.
\end{abstract}


\section{Introduction}
The popularity of sharing economy has reshaped urban mobility. Ride-sharing, in particular, is ubiquitous in large cities. As a consequence, numerous operational questions regarding ride-sharing have been raised. The operations research community has made significant efforts in addressing those questions by proposing a variety of routing and matching optimization models (see, e.g.,  \citep{tafreshian2020frontiers, martins2021optimizing, agatz2012optimization}). This short note examines a minimum fleet-size problem, which is a specific type of ride-sharing problem originally proposed by \citet{vazifeh2018addressing} and \citet{zhan2016graph}. The minimum fleet-size problem, as shown in those two papers, is effective in enhancing operational efficiency and reducing empty trip costs for ride-sharing companies, such as taxi services. In this paper, we study the minimum fleet-size problem 
and obtain a min-max theorem --- the minimum number of vehicles needed to cover all of the trips is equal to the maximum size of a set of trips that are pairwise incompatible.

\section{The Minimum Fleet Size Problem}
We begin by introducing the notation and formulation of the minimum fleet-size problem.
\paragraph{Basic definitions}
We are given a set $\mathcal{N}$ of trips to be covered. We define $\mathcal{N} := \{N_i = (p_i, T^p_{i}, d_i, T^d_{i}) : i \in [n]\}$; that is, the trip history shows that the $i$th trip picks up a passenger at time $T^p_{i}$ at point $p_i$ and drops off the fare at time $T^d_{i}$ at point $d_i$. In addition, we can estimate $time(s, t)$, the time that it takes 
to drive from location $s$ to location $t$.

The \textit{minimum fleet size problem} is defined as follows: given the set of trips $\mathcal{N}$, what is the minimum number of vehicles needed to handle all the trips?

\paragraph{Solving the minimum fleet-size problem via maximum bipartite matching}
As suggested in \citet{vazifeh2018addressing} and \citet{zhan2016graph}, we can solve the minimum fleet-size problem by reducing it to the bipartite matching problem and solving for the maximum matching.

Let $G = (D \cup P, E)$ be a bipartite graph where $D$ and $P$ are disjoint sets of vertices
and $E$ is the set of undirected edges, each of which has exactly one endpoint in $D$ and one endpoint in $P$. Let $D = \{(d_1, T^d_{1}),(d_2, T^d_{2}),\ldots,(d_n, T^d_{n})\}$ denote the set of drop-off nodes and $P = \{(p_1, T^p_{1}),(p_2, T^p_{2}),\ldots,(p_n, T^p_{n})\}$ denote the set of pickup nodes. The set $E$ is defined as follows: $$E = \{\{(d_i, T^d_{i}),(p_j, T^p_{j})\}: time(d_i, p_j) \leq T^p_{j} - T^d_{i}\};$$
in words, we include an edge exactly when the elapsed time between the drop-off time for $d_i$ and the pickup time for $p_j$, that is, $T_j^p - T_i^d$, is at least the time needed to travel between these two points $time(d_i,p_j)$. Hence, one vehicle can \textit{feasibly cover} both trip $N_j$ and $N_i$ since it can pick up a passenger in time at  $p_j$ after dropping off the previous fare at $d_i$.

We define a sequence of trips $N_{i_1}, N_{i_2},\ldots, N_{i_k}$ in ascending order of drop-off time to be a \textit{feasible trajectory} if one vehicle can feasibly cover each consecutive pair $(N_{i_s}, N_{i_t})$, where $t=s+1$. Observe that we can interpret any feasible trajectory $N_{i_1}, N_{i_2},\ldots, N_{i_k}$ as the matching consisting of the edges 
$$\{(d_{i_1}, T^d_{i_1}), (p_{i_2}, T^p_{i_2})\}, \{(d_{i_2}, T^d_{i_2}), (p_{i_3}, T^p_{i_3})\},\ldots, \{(d_{i_{k-1}}, T^d_{i_{k-1}}), (p_{i_k}, T^p_{i_k})\}.$$

For completeness, we include the proof of the correspondence between matchings and trajectories due to \citet{zhan2016graph} and \citet{vazifeh2018addressing}.

\begin{theorem} 
\label{matching}
Given any matching of size $m$, there exists a covering of all $n$ trips with $n-m$ vehicles.
\end{theorem}

\begin{proof}
Consider any matching $M$ in $G$ (that is, $M \subseteq E$) where $|M| = m$, and take any node $(d_{i_1}, T^d_{i_1})$ for which $(p_{i_1}, T^p_{i_1})$ is not an endpoint of any edge in $M$. There must be such a pick-up node $(p_j, T^p_{j})$ since the earliest pickup occurs at a time that is earlier than the earliest drop-off, and hence $(p_j, T^p_{j})$ cannot be the endpoint of any edge in $E$.

Starting with $(d_{i_1}, T^d_{i_1})$, we check if the current drop-off node is matched in $M$; suppose that $\{(d_{i_1}, T^d_{i_1}), (p_{i_2}, T^p_{i_2})\}$ is an edge in $M$ . This means that we can extend the trajectory from trip $N_{i_1}$ to trip $N_{i_2}$, and iteratively, we can continue in this way until we reach some drop-off node that is not matched. Since the number of $D$ nodes is equal to the number of $P$ nodes, and the number of matched nodes in each must always be equal, it follows that the number of unmatched nodes in $D$ must equal the number of unmatched nodes in $P$; since the latter is positive, so must the former. We can repeat this procedure, starting with any unmatched node in $P$, that is, any pickup point that is not paired with a corresponding earlier drop-off point - these are the starting points of feasible trajectories. Thus, the number of feasible trajectories that could cover all of the $n$ trips is equal to the total number of unmatched nodes in $P$, which is $n-m$. Since each feasible trajectory needs one vehicle, we find a one-to-one correspondence between a matching of size $m$ and a covering of all $n$ trips that uses $n-m$ vehicles.
\end{proof}

\section{The Min-max Theorem}

Suppose that $\textsc{P}_1$ denotes the problem of minimizing the total number of vehicles needed to cover all of the trips. Any two trips $N_i = (p_i, T^p_{i}, d_i, T^d_{i})$ and $N_j = (p_j, T^p_{j}, d_j, T^d_{j})$, $i \neq j \in [n]$, are considered {\it compatible} with each other if a vehicle can feasibly cover both, either by reaching from $d_i$ to $p_j$ or from $d_j$ to $p_i$ in time, that is, $$time(d_i, p_j) \leq T^p_{j} - T^d_{i}$$ or $$time(d_j, p_i) \leq T^p_{i} - T^d_{j}.$$ Furthermore, suppose that $\textsc{P}_2$ denotes the problem of maximizing the number of trips that are all pairwise incompatible (that is, no pair of them are compatible). Let $val(\textsc{P}_1)$ and $val(\textsc{P}_2)$ denote the optimal values of $\textsc{P}_1$ and $\textsc{P}_2$, respectively.

\begin{prop}\label{weak duality}
Suppose that there is a feasible solution for $\textsc{P}_2$ in which there are $k$ trips $(1\leq k\leq n)$, with the property that each pair of them is incompatible. Then, in any feasible solution to $\textsc{P}_1$ for this input, there must be at least $k$ vehicles used to cover all $n$ trips in $\mathcal{N}$.
\end{prop}

\begin{proof}
Let $I \subseteq \mathcal{N}$ be a feasible solution to $\textsc{P}_2$, where $|I| = k$; suppose that $I = \{(p_l, T^p_{l}, d_l, T^d_{l}): l \in K\}$ where $K \subseteq \{1,\ldots,n\}$ and $|K| = k$. 
Suppose for contradiction that $\mathcal{N}$ could be covered by fewer than $k$ vehicles; consequently, $I$ can be covered by fewer than $k$ vehicles. Hence, by the pigeonhole principle, there must exist a pair of trips in $I$ that is served by the same vehicle. In other words, for some $i \neq j \in K$, $time(d_i, p_j) \leq T^p_{j} - T^d_{i}$. This implies that there exists a compatible pair of trips in $I$, contradicting the fact that $I$ is a feasible solution to $\textsc{P}_2$.
\end{proof}

\begin{prop}\label{k pairs}
Consider two disjoint sets $P= \{p_1,\ldots,p_n\}$ and $D = \{ d_1,\ldots, d_n\}$, and a subset $I \subseteq P \cup D$, where $|I| = n + k$,
for some $k \geq 0$.  Then there exists a subset
$K \subseteq \{1,\ldots,n\}$ where $|K|=k$ and 
$S = \cup_{i \in K} \{ p_i, d_i\} \subseteq I$.
\end{prop}
\begin{proof}
We will prove this claim by induction on $n$. 

\smallskip
\noindent
{\bf Base case:} If $n=1$, then either $k=0$
or $k=1$; in the former case, 
we can choose $K = \emptyset$, and in the latter case, since $|I| = 2$, it 
follows that $I = \{p_1,d_1\}$, and so $K=\{1\}$ satisfies the desired property.

\smallskip
\noindent
{\bf Inductive step:} Now suppose that the claim holds for all inputs where 
$P$ and $D$ are of size $n$,
and consider an input where $|P|=|D| = n+1$, and $|I| = (n+1)+k$, where 
$k\geq 0$. Again, if $k=0$, there is nothing to prove, since we can take 
$K=\emptyset$. Now assume that $k \geq 1$. Since $|I| > n+1$, a pigeonhole
argument implies that there must exist some index $j$ such that $p_j \in I$ and
$d_j \in I$. Consider the case now where $P' = P - \{p_j\}$ and 
$D' = D - \{d_j\}$, and $I' = I - \{p_j, d_j\}$. Hence $|P'| =|D'| = n$ and 
$|I'| = n + 1 +k -2 = n + (k-1)$, where $k-1 \geq 0$. Hence, 
by the inductive hypothesis,
there must exist a set $K'$ of cardinality $k-1$, such that the set
$S' = \cup_{i \in K'} \{ p_i, d_i \}$ is a subset of $I'$. But then
$K = K' \cup \{ j \}$ is a valid choice for $P$ and $D$ of cardinality $k$ for
which $\cup_{i \in K} \{p_i,d_i\} \subseteq I$.

\smallskip
\noindent
This proves the proposition.
\end{proof}

\begin{theorem}
\label{min_max_theorem}
The maximum size of a set of trips that are pairwise incompatible is equal to the minimum number of vehicles needed to cover all of the trips; or equivalently, $val(\textsc{P}_1) = val(\textsc{P}_2)$.
\end{theorem}

\begin{proof}
By Proposition \ref{weak duality}, we know that $val(\textsc{P}_1) \leq val(\textsc{P}_2)$. Thus, it is sufficient to show that, given an optimal solution to $\textsc{P}_1$ of cardinality $k$, there is a feasible solution to $\textsc{P}_2$ of cardinality $k$.

We have already seen in Theorem \ref{matching} that there is a one-to-one correspondence between matchings of size $m$ in the graph $G=(D,P,E)$, where $|D| = |P|=n$, and coverings of all trips that use $n-m$ vehicles, or equivalently, between coverings with $k$ vehicles and matchings of size $n-k$. Thus, if $k$ is the minimum number of vehicles that suffice, then $n-k$ is the size of a maximum matching in $G$.

We want to show that there is a set of $k$ pairwise incompatible trips.
Let $m=n-k$. By König's theorem, a maximum matching of size $m$ implies that there is a vertex cover $C$ of size $m$, where a vertex cover is a set of vertices that contains at least one endpoint of each edge in the graph. The complement of any vertex cover is an independent set $I$ in $G$; that is, $I= P \cup D - C$ is a set of pairwise non-adjacent vertices.  Furthermore, $|I|=2n-m = n+k$. 

By Proposition \ref{k pairs}, there is a subset $S \subseteq I$ of the form $S = \cup_{j \in K} \{(p_j,T_j^p),(d_j,T_j^d)\}$ for some $K \subseteq \{1,\ldots,n\}$ and $|K| = k$. Since $I$ is an independent set in $G$, so is $S$. Consequently, $K$ indexes a set of pairwise incompatible trips of size $k$.
\end{proof}

\section{Future Work}
A natural extension to this paper is a min-max theorem on the full formulation of the minimum fleet-size problem mentioned in \citet{vazifeh2018addressing} and \citet{zhan2016graph}. The full formulation includes an additional upper-bound constraint on the allowed waiting time for the driver at the pickup location. Let $\delta$ denote this bound, and so, in order for a vehicle to cover two trips $N_i$ and $N_j$ (assuming that $T_j^p > T_i^d$), we need
the elapsed time between the drop-off time at $d_i$ and the pickup time at $p_j$, $T_j^p - T_i^d$, to be at most $time(d_i, p_j) + \delta$ in addition to the arrival-in-time requirement of $T_j^p - T_i^d \geq time(d_i, p_j)$. We do not know of an analogous result to Theorem \ref{min_max_theorem} in this more general setting.

\section{Acknowledgements}
This work is partially funded by Cornell ELI Undergraduate Research Funds.

\section{Declaration of Interests}
The authors declare that they have no known competing financial interests or personal relationships that could have appeared to influence the work reported in this paper.

\bibliographystyle{plainnat}

\bibliography{refs}

\begin{thebibliography}{5}
\providecommand{\natexlab}[1]{#1}
\providecommand{\url}[1]{\texttt{#1}}
\expandafter\ifx\csname urlstyle\endcsname\relax
  \providecommand{\doi}[1]{doi: #1}\else
  \providecommand{\doi}{doi: \begingroup \urlstyle{rm}\Url}\fi

\bibitem[Agatz et~al.(2012)Agatz, Erera, Savelsbergh, and
  Wang]{agatz2012optimization}
Niels Agatz, Alan Erera, Martin Savelsbergh, and Xing Wang.
\newblock Optimization for dynamic ride-sharing: A review.
\newblock \emph{European Journal of Operational Research}, 223\penalty0
  (2):\penalty0 295--303, 2012.

\bibitem[Martins et~al.(2021)Martins, de~la Torre, Corlu, Juan, and
  Masmoudi]{martins2021optimizing}
Leandro do~C Martins, Rocio de~la Torre, Canan~G Corlu, Angel~A Juan, and
  Mohamed~A Masmoudi.
\newblock Optimizing ride-sharing operations in smart sustainable cities:
  Challenges and the need for agile algorithms.
\newblock \emph{Computers \& Industrial Engineering}, 153:\penalty0 107080,
  2021.

\bibitem[Tafreshian et~al.(2020)Tafreshian, Masoud, and
  Yin]{tafreshian2020frontiers}
Amirmahdi Tafreshian, Neda Masoud, and Yafeng Yin.
\newblock Frontiers in service science: Ride matching for peer-to-peer ride
  sharing: A review and future directions.
\newblock \emph{Service Science}, 12\penalty0 (2-3):\penalty0 44--60, 2020.

\bibitem[Vazifeh et~al.(2018)Vazifeh, Santi, Resta, Strogatz, and
  Ratti]{vazifeh2018addressing}
Mohammad~M Vazifeh, Paolo Santi, Giovanni Resta, Steven~H Strogatz, and Carlo
  Ratti.
\newblock Addressing the minimum fleet problem in on-demand urban mobility.
\newblock \emph{Nature}, 557\penalty0 (7706):\penalty0 534--538, 2018.

\bibitem[Zhan et~al.(2016)Zhan, Qian, and Ukkusuri]{zhan2016graph}
Xianyuan Zhan, Xinwu Qian, and Satish~V Ukkusuri.
\newblock A graph-based approach to measuring the efficiency of an urban taxi
  service system.
\newblock \emph{IEEE Transactions on Intelligent Transportation Systems},
  17\penalty0 (9):\penalty0 2479--2489, 2016.

\end{thebibliography}

\end{document}